\documentclass[conference]{IEEEtran}
\IEEEoverridecommandlockouts
\usepackage{cite}
\usepackage{amsmath,amssymb,amsfonts}
\usepackage{algorithmic}
\usepackage{graphicx}
\usepackage{textcomp}
\usepackage{xcolor}
\usepackage{booktabs}
\usepackage{float}
\def\BibTeX{{\rm B\kern-.05em{\sc i\kern-.025em b}\kern-.08em
    T\kern-.1667em\lower.7ex\hbox{E}\kern-.125emX}}
\begin{document}

\title{Query Implied Generative Engine Optimization\\}

\author{\IEEEauthorblockN{Shilpa Ramakrishna}
\IEEEauthorblockA{\textit{Department of Computer Science} \\
\textit{San Jose State University}\\
San Jose, CA, USA \\
shilpa.ramakrishna@sjsu.edu}
\and

\IEEEauthorblockN{William B. Andreopoulos}
\IEEEauthorblockA{\textit{Department of Computer Science} \\
\textit{San Jose State University}\\
San Jose, CA, USA \\
william.andreopoulos@sjsu.edu}

}

\maketitle

\begin{abstract}
The landscape of search has changed drastically with how people look for information online. Traditional search engines are being replaced by Generative Search Engines (GSEs), which use Large Language Models (LLMs) to generate natural language responses to user queries. For content creators, visibility is no longer solely determined by ranking in search results but by being cited within generated responses. But Generative Search Engines are black-boxes, leading to the emergence of Generative Engine Optimization (GEO), a set of techniques aimed at improving content visibility in generative search settings. Most existing approaches rely on the explicit queries or query derived signals to align content to better suit user needs. We propose Query Implied Generative Engine Optimization (QI-GEO) to infers user intent directly from the document. Our approach approximates document's intent space and identifies content that may be missing yet relevant to answer potential user queries. Evaluation on GEO-Bench and Extended GEO-Bench demonstrated improvements across objective and subjective metrics. QI-GEO improved objective scores by up to 15.9\% and subjective scores by up to 17.6\%, while yielding nearly twice as many citation gains as citation losses. These results suggest that document-derived approximations of user intents can improve visibility without relying on explicit query inputs.
\end{abstract}


\section{Introduction}
For more than twenty years, search engines such as Google, Yahoo, and others have been the primary means for people to find information online. These systems relied on keyword-based retrieval and returned ranked lists of websites known as Search Engine Results Pages (SERPs). Content creators used Search Engine Optimization (SEO) techniques to improve rankings and increase visibility within these results. With the introduction of Large Language Models (LLMs), a new type of search system emerged: Generative Search Engines (GSEs) \cite{sge}. Unlike traditional search engines, GSEs generate natural language answers by retrieving relevant sources and synthesizing information into a coherent response \cite{rag}. These responses often include citations to the sources used during generation \cite{verifiability}. As users increasingly interact with search systems through complete questions rather than keywords, visibility now depends not only on ranking in search results but also on being selected and cited within generated answers \cite{seo_vs_geo}. As a result, traditional SEO practices alone may be insufficient to maintain visibility.

To address visibility in generative search environments, the concept of Generative Engine Optimization (GEO) was introduced \cite{geo}. GEO refers to a set of techniques designed to increase the likelihood that a document will be cited in responses generated by GSEs. The foundational GEO work introduced the concept of visibility in generative outputs and proposed objective and subjective impression metrics for evaluating it \cite{geo}. It also demonstrated that even simple content modifications can significantly affect visibility. Subsequent approaches such as RAID \cite{raid}, C-SEO \cite{cseo_bench}, and IF-GEO \cite{ifgeo} further explored GEO and proposed alternative optimization strategies. However, most existing methods rely on explicit queries or query-derived signals to guide optimization. In practice, content creators rarely know the full range of questions users may ask about a topic, making query-dependent optimization difficult in black-box generative search systems.

Queries are often used as a proxy for user intent. In this work, we propose Query-Implied Optimization (QI-GEO), a framework that seeks to infer user intent directly from the document rather than relying on explicit query inputs. The key idea behind QI-GEO is that documents contain entities and relationships that capture their underlying semantics and reflect the information users may be interested in. By representing these relationships in a structured form and expanding them using external knowledge, QI-GEO approximates the broader semantic space surrounding a document without requiring query generation.

During this expansion process, concepts emerge that are closely related to the document’s topic but are not explicitly covered in its content. These concepts can be viewed as potential informational gaps that may limit a document’s ability to address diverse user information needs. QI-GEO leverages these gaps as opportunities for optimization. By identifying and incorporating relevant missing information, QI-GEO improves content coverage while grounding optimization in the document’s semantic structure rather than generated queries. This provides a more stable representation of user intent and helps improve visibility across unseen queries.

In summary, the contributions of this work are as follows. First, we propose Query-Implied Optimization (QI-GEO), a GEO framework that infers latent user intent directly from document semantics without relying on explicit query inputs. Second, we evaluate QI-GEO on GEO-Bench and ExtendedGEOBench using objective, subjective, citation-based, and stability metrics, demonstrating consistent improvements across single-query and multi-query evaluation settings.

\section{Related Works}

Most GEO research assumes a fixed-retrieval setting, where the candidate document pool associated with a query remains constant throughout evaluation \cite{geo, raid, ifgeo}. This setup isolates the effect of content modifications from retrieval variability and has become the standard evaluation protocol for comparing GEO methods \cite{geo, raid, ifgeo}.

\subsection{Generative Engine Optimization}

Generative Engine Optimization (GEO) was introduced as a set of techniques for improving document visibility in generated responses \cite{geo}. The foundational GEO work proposed objective and subjective visibility metrics and demonstrated that content modifications such as adding citations, statistics, and authoritative references can significantly improve visibility. Subsequent works such as RAID \cite{raid}, IF-GEO \cite{ifgeo}, and C-SEO \cite{cseo_bench} extended GEO through alternative optimization strategies and evaluation frameworks.

GEO methods have evolved from rule-based rewriting to intent-aware and multi-query optimization. RAID uses intent inference and structured rewriting plans to better align content with user information needs \cite{raid}, while IF-GEO focuses on optimizing documents across multiple related queries through a diverge-then-converge framework \cite{ifgeo}. Despite these differences, most GEO approaches rely on explicit queries or query-derived signals to guide optimization.

\subsection{Query and Intent Modeling}

User intent is a central component of GEO because visibility depends on how well a document aligns with user information needs. Early GEO approaches optimize content with respect to known queries \cite{geo}. More recent methods generate related queries or query variants to approximate a broader intent space \cite{raid, ifgeo}.

RAID further models intent through user-role simulation and reflection mechanisms \cite{raid}, while IF-GEO employs reverse retrieval to generate potential queries for a document \cite{ifgeo}. Although these methods differ in how intent is represented, they generally treat queries as the primary mechanism for approximating user needs.

\subsection{Evaluation Frameworks and Metrics}

GEO-Bench is the most widely used benchmark for evaluating GEO methods and contains 10,000 queries spanning diverse domains and query sources \cite{geo}. RAID extends this benchmark through ExtendedGEOBench, which groups semantically related queries to support robustness and multi-query evaluation \cite{raid}. IF-GEO further uses this setting to study optimization consistency across related queries \cite{ifgeo}.

Existing work evaluates visibility using objective metrics such as Word Count and Position-Adjusted Word Count (PAWC) \cite{geo}, subjective metrics based on LLM-as-a-judge approaches such as G-EVAL \cite{geval}, and stability metrics such as Worst-Case Performance, Downside Risk, and Win-Tie Rate \cite{ifgeo}.

Existing GEO methods have progressed from rule-based rewriting to intent-aware and multi-query optimization. However, most approaches model user intent through explicit queries, generated queries, or query-derived signals. Since content creators rarely know the complete space of possible user queries, there remains a need for methods that can infer intent directly from the document itself. This gap motivates the document-centric approach explored in this work.

\section{Problem Formulation}

A Generative Search Engine (GSE) responds to a query $q$ by retrieving a set of candidate documents $C_q$ and generating a natural language response $r_q$ from the retrieved evidence \cite{geo, rag}:

\begin{equation}
r_q = \text{Generate}(q, C_q)
\end{equation}

Unlike traditional search engines, visibility in GSEs depends on whether a document is incorporated and cited in the generated response rather than its ranking position \cite{geo}. GEO methods seek to improve this visibility through document optimization. Let $v(D,q)$ denote the visibility of document $D$ for query $q$, measured using objective or subjective visibility metrics \cite{geo, geval}. Following prior GEO work, visibility may be measured using objective metrics based on citation contribution and placement, or subjective metrics that capture the perceived influence of a document in the generated response \cite{geo, geval}.

Given a document $D$, a GEO method $M$ produces an optimized document $D' = M(D)$. For a query $q_i$, the visibility gain is defined as:

\begin{equation}
\Delta v_i = v(D', q_i) - v(D, q_i)
\end{equation}

Since a document is expected to perform across multiple related queries, the overall optimization effect is measured by aggregating visibility gains over a query set $Q(D)$:

\begin{equation}
\Delta = A({\Delta v_i}{i=1}^{m})
\end{equation}

where $A(\cdot)$ is an aggregation function such as the mean or minimum visibility gain.

Existing GEO methods assume that relevant queries are either known or can be approximated through query generation \cite{geo, raid, ifgeo}. However, content creators rarely have access to the full set of queries that may retrieve a document. Let $P(D)$ denote the unknown distribution of queries associated with document $D$. The optimization objective is therefore to maximize the expected visibility gain:

\begin{equation}
\max_M \mathbb{E}{q \sim P(D)}
\left[
v(M(D),q)-v(D,q)
\right]
\end{equation}

where $q \sim P(D)$ denotes a query sampled from the unknown retrieval scenario of $D$. Existing methods approximate user intent through known queries, generated queries, or query-derived signals \cite{geo, raid, ifgeo}. Since $P(D)$ is unavailable at optimization time, the challenge is to derive an alternative representation of user intent directly from the document itself. The next section presents QI-GEO, a document-centric approach that addresses this problem through semantic expansion.

\section{Methodology}

\begin{figure*}[t]
\centering
\includegraphics[width=\textwidth]{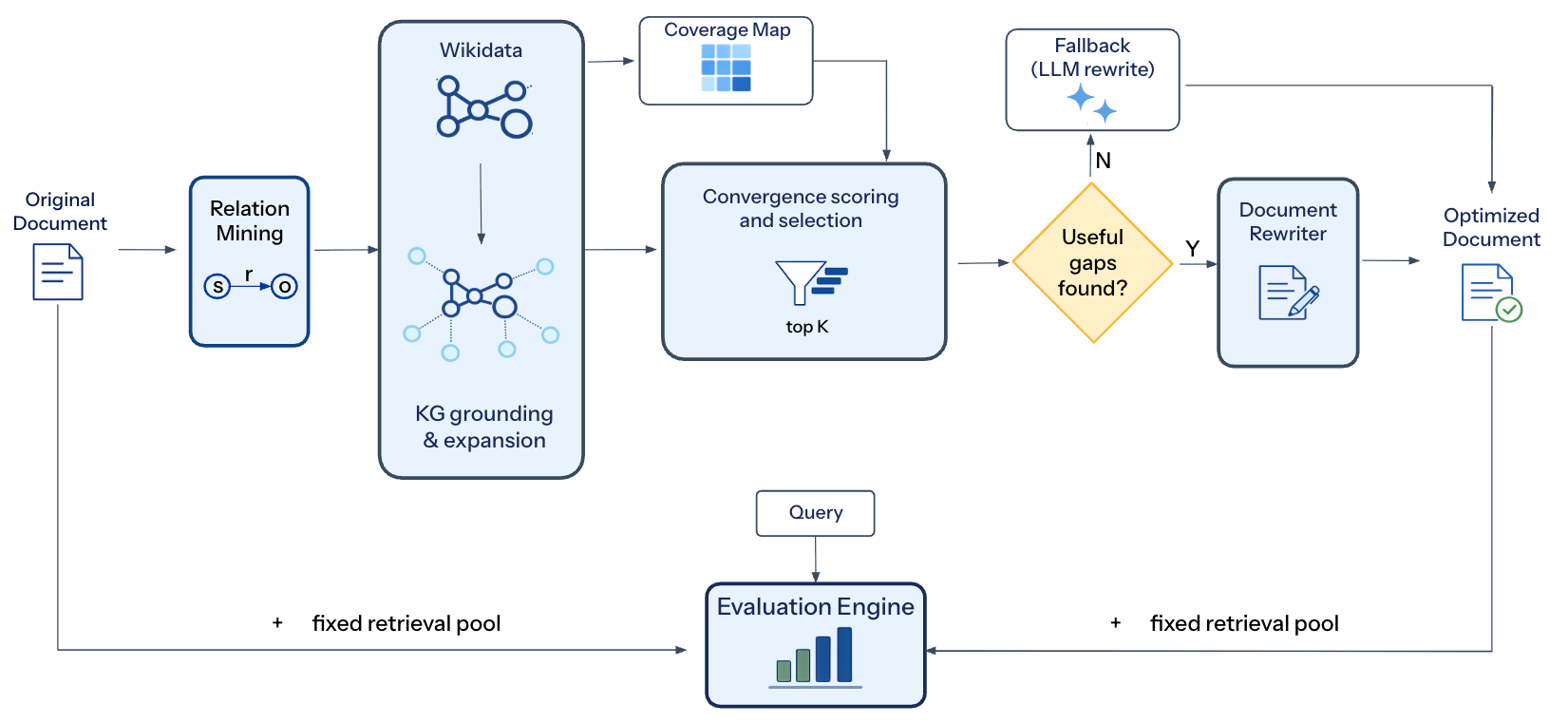}
\caption{Overview of the Query-Implied Optimization (QI-GEO) framework.}
\label{fig:qio-pipeline}
\end{figure*}

\subsection{Overview of QI-GEO}

Query-Implied Optimization (QI-GEO) is a document-centric GEO framework that improves visibility in generative search engines without relying on explicit query inputs. Instead of optimizing content for known or generated queries, QI-GEO attempts to infer user intent directly from the document’s semantic structure. The central idea is that entities and relationships present in a document reflect the information needs associated with its topic. By expanding these relationships using external knowledge, QI-GEO approximates the broader semantic space surrounding the document and identifies concepts that are relevant but not explicitly covered.

Figure~\ref{fig:qio-pipeline} illustrates the overall workflow of QI-GEO. The framework first extracts semantic relationships from the document and grounds them in a knowledge graph. The resulting graph is then expanded to discover related concepts and construct a coverage map of the document’s existing semantic content. Candidate informational gaps are ranked according to their relevance to the document’s topic, and the most promising gaps are converted into targeted edits. These edits are incorporated into the document to improve coverage of potential user information needs while preserving the original content and structure.

\subsection{Relation Mining and Knowledge Graph Construction}

QI-GEO begins by converting the document into a structured semantic representation that can be expanded and analyzed. Since the framework models intent through entities and their relationships, the first step is to identify the key concepts discussed in the document and how they are connected.

To reduce ambiguity, coreference resolution is first applied to the entire document so that pronouns and other referring expressions are replaced with their corresponding entities. This preserves semantic relationships that span multiple sentences and prevents incomplete relations from being extracted. The resolved document is then processed using a relation extraction model, which converts text into a collection of subject–relation–object triplets. Each triplet captures a semantic relationship expressed in the document and serves as the interface between unstructured text and structured knowledge. Triplets are used because they provide a compact representation of document semantics while remaining compatible with knowledge graph entities and relations.

The extracted entities and relations are subsequently linked to identifiers in a knowledge graph. Entity linking grounds document concepts in a shared semantic space and provides a consistent representation for downstream expansion. The resulting grounded triplets collectively form a document knowledge graph that serves as the semantic foundation for the remainder of the QI-GEO pipeline.

\subsection{Knowledge Expansion and Coverage Mapping}

Once the document graph has been constructed, QI-GEO expands it to explore information related to the document’s topic. The objective of this stage is to approximate the broader semantic space surrounding the document and identify concepts that may be relevant to potential user information needs. Rather than generating possible queries, QI-GEO explores concepts that are semantically connected to the document through a knowledge graph.

For each linked entity, QI-GEO retrieves neighboring concepts using one-hop expansion. Restricting expansion to immediate neighbors keeps the process focused on concepts that are directly related to the document and reduces the introduction of unrelated information. This also makes the resulting gaps easier to justify, since every candidate remains only one semantic step away from concepts already present in the document. To further improve relevance, relation types that primarily encode metadata or navigational information are filtered, while relations that capture definitional, causal, or explanatory information are prioritized.

Alongside expansion, QI-GEO builds a coverage map that records the entity–relation pairs already expressed in the document. The purpose of the coverage map is to distinguish genuinely missing concepts from information that is already present. Concepts discovered during expansion that are not represented in the coverage map are treated as candidate informational gaps. These candidates are passed to the next stage, where they are evaluated and ranked according to their relevance to the document.

\subsection{Gap Discovery and Ranking}

Knowledge graph expansion typically produces a large number of candidate concepts, many of which are only weakly related to the document. The objective of this stage is to identify concepts that are both strongly implied by the document and genuinely absent from its content.

QI-GEO ranks candidate concepts using a convergence-based scoring strategy. The intuition is that concepts independently supported by multiple entities in the document are more likely to represent meaningful informational gaps than concepts originating from a single source. A candidate concept that is repeatedly implied by different parts of the document provides stronger evidence of an underlying information need and is therefore assigned higher importance.

To prevent semantically similar relations from fragmenting evidence, relation properties are first grouped according to their semantic function. Examples include definitional, causal, mechanistic, and metadata-oriented relations. Each relation group is assigned a facet weight that reflects its usefulness for reader-facing content. Definitional and explanatory relations are generally assigned higher weights than navigational or metadata relations.

The structural score of a candidate concept is computed as:

\begin{equation}
S_{\text{struct}} =
\text{conv}(g,d)\times
w_{\text{facet}}(g)\times
c_{\text{src}}(e)
\end{equation}

where $\text{conv}(g,d)$ measures the degree of convergence within the document, $w_{\text{facet}}(g)$ is the weight assigned to relation group $g$, and $c_{\text{src}}(e)$ measures the importance of the source entity. Convergence is defined as the number of distinct document entities that independently support a candidate relation group:

\begin{equation}
\text{conv}(g,d)=
\left|
{e \in E_d :
\exists \text{ candidate } (g,n)
\text{ from } e
}
\right|
\end{equation}

Source entity importance is estimated using normalized mention frequency:

\begin{equation}
c_{\text{src}}(e)=
\frac{\text{count}(e,d)}
{\max_{e' \in E_d}\text{count}(e',d)}
\end{equation}

This gives greater influence to concepts originating from entities that are central to the document while reducing the effect of isolated mentions.

After scoring, candidate concepts are evaluated against the coverage map. Exact coverage checks remove concepts whose entity--relation pairs already exist in the document. A second semantic coverage check compares candidate concepts against document content to identify information that is already expressed using different wording. Candidates that pass both checks are treated as true informational gaps.

The remaining gaps are ranked according to their scores, and the highest-ranked candidates are selected for document rewriting.

\subsection{Document Rewriting}

The final stage of QI-GEO converts the selected informational gaps into targeted document revisions. Rather than performing unrestricted rewriting, QI-GEO generates structured edit specifications that identify what information should be added and where it should be incorporated within the document.

Each informational gap is transformed into an edit specification consisting of three components: an anchor, a suggestion, and a necessity score. The anchor identifies the sentence or passage most closely related to the gap, the suggestion describes the missing concept that should be incorporated, and the necessity score reflects the importance of addressing the gap. Anchoring edits to existing content provides a clear location for modification and helps maintain consistency with the surrounding context.

Edits are applied sequentially using the current version of the document. To preserve document structure and minimize unintended changes, the rewriter is instructed to make localized additions rather than perform large-scale rewrites. In most cases, edits are implemented by appending a short clause or sentence to the selected anchor while leaving the remainder of the document unchanged.

The use of anchored edits improves traceability throughout the optimization process. Every revision can be linked back to a specific informational gap and a corresponding location in the original document, making the resulting changes easier to inspect and verify. 

The final output is an optimized document that incorporates high-value missing information while preserving the intent, structure, and content of the original text.

\section{Experimental Results}

\subsection{Research Questions}

The experiments are designed to evaluate whether document semantics can be used as an alternative to query-based optimization in generative search engines. Specifically, we investigate the following research questions:

\begin{itemize}
    \item \textbf{RQ1:} Can user intent be inferred directly from a document without access to explicit user queries?

    \item \textbf{RQ2:} Can semantic expansion reveal missing but relevant information that improves document visibility in generative search engines?

    \item \textbf{RQ3:} Do the visibility gains achieved through query-independent optimization remain consistent across different formulations of the same information need?
\end{itemize}

\subsection{Datasets and Benchmarks}

Experiments are conducted using GEO-Bench \cite{geo} and ExtendedGEOBench \cite{raid}. GEO-Bench is the standard benchmark for evaluating visibility in generative search engines and contains 10,000 query-document instances spanning a diverse set of domains. Each query is associated with a fixed retrieval pool consisting of five candidate documents and a designated target document for optimization.

To evaluate robustness under query variation, we additionally use ExtendedGEOBench \cite{raid}, which groups semantically related queries that express the same information need through different phrasings. This benchmark makes it possible to evaluate whether an optimization method remains effective across multiple formulations of a topic rather than a single query.

Following prior GEO work \cite{geo, raid, ifgeo}, all experiments use a fixed-retrieval setting where the candidate document pool remains unchanged throughout evaluation. This isolates the effect of document modifications from retrieval variability and ensures that changes in visibility can be attributed solely to the optimization process.

For each query, the target document provided by the benchmark is optimized using QI-GEO and evaluated against its original version while keeping the remaining documents in the retrieval pool unchanged.

\subsection{Experimental Configuration}

QI-GEO is instantiated using a relation extraction, knowledge graph expansion, and LLM-based rewriting pipeline. Since the quality of extracted relations directly affects downstream expansion and gap discovery, we evaluate multiple relation extractors including REBEL \cite{rebel}, RELiK \cite{relik}, spaCy \cite{spacy}, and an LLM-based extractor built using Qwen2.5 \cite{qwen25}. REBEL is used for all reported results because it consistently produced the most useful triplets for knowledge graph expansion. Its relation vocabulary is naturally aligned with knowledge graph properties, resulting in more reliable entity linking and downstream expansion.

Extracted triplets are linked to a knowledge graph and expanded using one-hop neighborhood retrieval. To keep expansion focused on concepts that are directly related to the document, generic and metadata-oriented relations are filtered while definitional, explanatory, and causal relations are retained. Candidate concepts are then ranked using the convergence-based scoring strategy described in Section~IV and filtered using semantic similarity and coverage checks.

Selected informational gaps are converted into edit requests and incorporated into the document using Qwen2.5 7B \cite{qwen25}. Rewriting is performed through localized edits anchored to existing document content so that missing information can be introduced without substantially altering the original structure or meaning of the document.

\subsection{Evaluation Metrics}

We evaluate QI-GEO using objective, subjective, and stability metrics that are commonly used in GEO research.

\textbf{Objective metrics:} We report Word Count Impression (WC), Position Impression (PI), and Position-Adjusted Word Count (PAWC) \cite{geo}. WC measures the proportion of generated content attributed to a document, while PI measures how prominently that content appears based on citation position. PAWC combines these two signals by weighting word contributions according to their position in the response, assigning greater importance to citations that appear earlier.

\begin{equation}
\text{Imp}{pwc}(c_i, r) =
\frac{\sum{s \in S_{c_i}} |s| \cdot e^{-\frac{\text{pos}(s)}{|S|}}}
{\sum_{s \in S_r} |s|}
\end{equation}

where $S_{c_i}$ is the set of sentences citing document $c_i$ and $S_r$ is the set of all sentences in the response. If a sentence contains multiple citations, the contribution is divided equally among the cited sources. Since PAWC captures both citation volume and citation prominence, it is used as the primary objective metric throughout this work.

\textbf{Subjective metrics:} We also report subjective impression scores following the GEO evaluation framework \cite{geo}. These scores estimate how visible and influential a document appears from a user perspective. The evaluation considers seven dimensions: relevance, influence, uniqueness, diversity, click follow-up likelihood, subjective position, and subjective volume. Each dimension is scored on a five-point scale, and the final subjective score is computed as the average across all dimensions.

\textbf{Stability metrics:} To evaluate robustness across query variations, we report Mean Gain, Win-Tie Rate (WTR), Worst-Case Performance (WCP), and Downside Risk (DR) \cite{ifgeo}. Mean Gain measures the average relative improvement over the original document. WTR measures the fraction of queries for which the optimized document performs at least as well as the original. WCP reports the largest observed performance drop across all evaluated queries, while DR measures the magnitude of negative outcomes. Together, these metrics provide a more complete picture of optimization reliability beyond average performance alone.

\section{Results and Discussion}

This section presents the experimental results of QI-GEO on GEO-Bench and ExtendedGEOBench. We evaluate whether query-independent optimization improves document visibility in generative search engines, whether semantic expansion identifies information that contributes to visibility gains, and whether these improvements remain robust across query variations. Unless otherwise stated, results compare the original document against its QIO-optimized version while keeping the retrieval pool and evaluation settings unchanged.

\subsection{Citation Coverage}

QI-GEO increases citation coverage in both the single-query and multi-query evaluation settings. Table~\ref{tab:cited_rate} reports the fraction of queries for which the target document is cited in the generated response.

\begin{table}[h]
\centering
\begin{tabular}{lcccc}
\toprule
\textbf{Setting} & \textbf{Original} & \textbf{QI-GEO} & \textbf{Gain} & \textbf{Gain \%} \\
\midrule
Single Query & 0.657 & 0.726 & +0.069 & +10.5\% \\
Multi-Query $\times$5 & 0.713 & 0.772 & +0.059 & +8.3\% \\
\bottomrule
\end{tabular}
\vspace{2mm}
\caption{Cited rate comparison}
\label{tab:cited_rate}
\end{table}

In the single-query setting, citation coverage increases from 0.657 to 0.726, corresponding to a relative improvement of 10.5\%. This means that roughly one in ten documents that were previously ignored by the generator become cited after optimization. Similar improvements are observed in the multi-query setting, where citation coverage increases by 8.3\%.

\begin{figure}[H]
\centering
\includegraphics[width=0.7\linewidth]{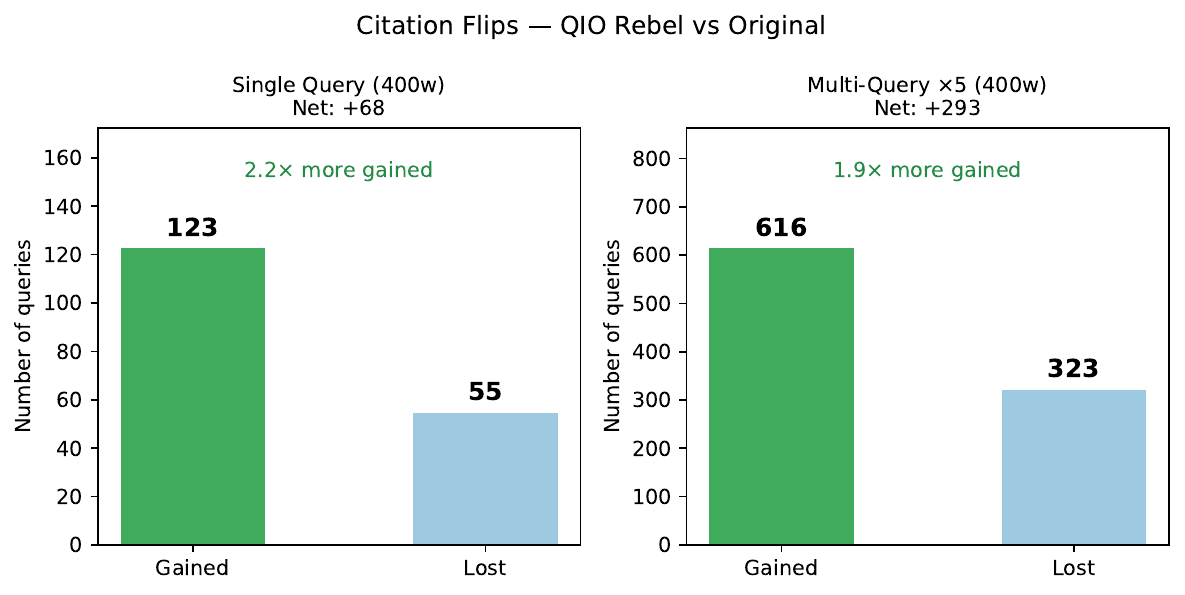}
\caption{Citation Flips: Queries Gained vs Lost}
\label{fig:qio_citation}
\end{figure}

The gains suggest that QI-GEO makes relevant information easier for the generator to identify and incorporate into its responses. While the improvement is slightly smaller in the multi-query setting, the overall trend remains consistent across both evaluation scenarios.

Figure~\ref{fig:qio_citation} shows the number of queries that gained or lost citations after optimization. Across both settings, QI-GEO gains citations substantially more often than it loses them. In the single-query setting, citation gains outnumber losses by more than 2:1, while a similar pattern is observed in the multi-query setting. This indicates that the increase in citation coverage is not driven by a small number of large improvements, but by a consistent tendency for optimized documents to be cited more frequently than their original counterparts.

\subsection{Source Visibility}

Beyond being cited, a document must also contribute meaningfully to the generated response. We therefore evaluate visibility using Position-Adjusted Word Count (PAWC), which combines citation coverage and citation position into a single metric. Table~\ref{tab:wordpos_mean} reports the average PAWC scores before and after optimization.

\begin{table}[h]
\centering
\begin{tabular}{lcccc}
\toprule
\textbf{Setting} & \textbf{Original} & \textbf{QI-GEO} & \textbf{Gain} & \textbf{Gain \%} \\
\midrule
Single Query & 0.2137 & 0.2407 & +0.0270 & +12.6\% \\
Multi-Query $\times$5 & 0.2038 & 0.2363 & +0.0325 & +15.9\% \\
\bottomrule
\end{tabular}
\vspace{2mm}
\caption{Mean wordpos for original and QI-GEO rewritten documents.}
\label{tab:wordpos_mean}
\end{table}

QI-GEO improves mean PAWC from 0.2137 to 0.2407 in the single-query setting, corresponding to a relative improvement of 12.6\%. Similar gains are observed in the multi-query setting, where PAWC increases by 15.9\%. These results indicate that optimized documents are not only cited more frequently, but also tend to appear earlier in generated responses and contribute a larger share of the cited content.

While QI-GEO does not improve every query, the overall distribution is shifted toward positive gains. The average improvement remains positive in both evaluation settings, indicating that the gains are not driven by a small number of outliers.

To better understand consistency across queries, we additionally report Win-Tie Rate (WTR) and Downside Risk (DR). QI-GEO achieves a WTR of 0.70, meaning that the optimized document performs at least as well as the original document for approximately seven out of ten queries. The downside risk remains low at 0.011, suggesting that when QI-GEO does reduce visibility, the magnitude of the loss is generally small.

Taken together, these results suggest that QI-GEO provides consistent visibility improvements across a large portion of the benchmark while introducing relatively limited risk when optimization is unsuccessful.

\subsection{Subjective Evaluation}

To complement the objective metrics, we evaluate QI-GEO using subjective impression scores. These scores are generated using an LLM judge that evaluates each response across seven dimensions: relevance, influence, uniqueness, diversity, click follow-up likelihood, subjective position, and subjective volume. Table~\ref{tab:subj_mean} reports the average subjective scores before and after optimization.

\begin{table}[h]
\centering
\begin{tabular}{lcccc}
\toprule
\textbf{Setting} & \textbf{Original} & \textbf{QI-GEO} & \textbf{Gain} & \textbf{Gain \%} \\
\midrule
Single Query & 2.332 & 2.743 & +0.411 & +17.6\% \\
Multi-Query $\times$5 & 2.137 & 2.503 & +0.366 & +17.1\% \\
\bottomrule
\end{tabular}
\vspace{2mm}
\caption{Mean subjective score for original and QI-GEO-rewritten documents.}
\label{tab:subj_mean}
\end{table}

QI-GEO improves the mean subjective score from 2.332 to 2.743 in the single-query setting and from 2.137 to 2.503 in the multi-query setting, corresponding to relative improvements of 17.6\% and 17.1\%, respectively. The consistency of these gains across both evaluation settings suggests that the visibility improvements observed in the objective metrics are also reflected in the perceived usefulness and prominence of the document within generated responses.

\begin{figure}[H]
\centering
\includegraphics[width=0.8\linewidth]{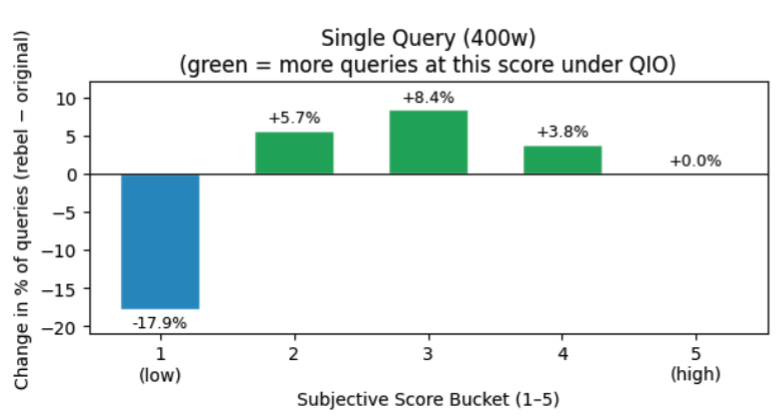}
\caption{Subjective Score Redistribution Shift - Single Query}
\label{fig:qio_subj1}
\end{figure}

\begin{figure}[H]
\centering
\includegraphics[width=0.8\linewidth]{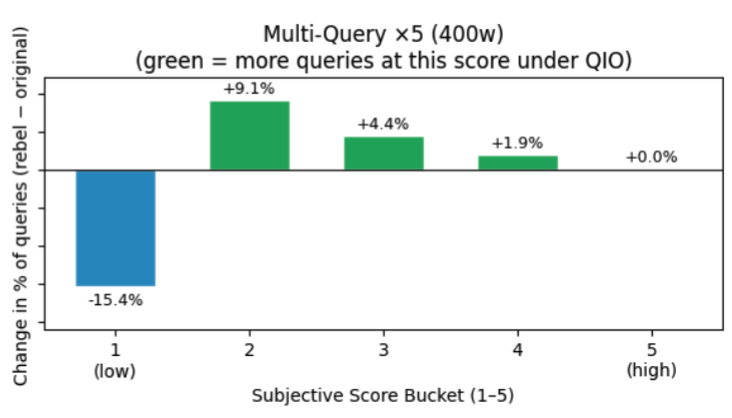}
\caption{Subjective Score Redistribution Shift - Multi Query}
\label{fig:qio_subj2}
\end{figure}

Figure~\ref{fig:qio_subj1} illustrates how the distribution of subjective scores changes after optimization. The largest reduction occurs at the lowest score range, while higher score ranges become more populated. This indicates that QI-GEO primarily improves documents that initially receive weak visibility, shifting them toward moderate and high visibility categories.

We additionally compute stability metrics for the subjective scores. QI-GEO achieves a WTR of 0.81, meaning that the optimized document receives an equal or higher subjective score than the original document for more than four out of five queries. Although some negative outcomes remain, the overall distribution is strongly skewed toward positive gains.

These results suggest that the improvements introduced by QI-GEO are not limited to citation statistics alone. The optimized documents are also perceived as more relevant, influential, and visible within generated answers.

\subsection{Multi-Query Robustness}

The previous experiments evaluate QI-GEO using a single query for each document. In practice, however, the same information need can be expressed through many different query formulations. To evaluate whether QI-GEO remains effective under query variation, we use ExtendedGEOBench \cite{raid}, which provides five semantically related query variants for each document.

Rather than analyzing each query independently, we aggregate results at the document level and evaluate whether the optimized document remains beneficial across multiple formulations of the same information need. Table~\ref{tab:mq_stability} summarizes the resulting robustness metrics.

\begin{table}[h]
\centering
\begin{tabular}{lc}
\toprule
\textbf{Metric} & \textbf{Value} \\
\midrule
Docs with net-positive mean gain across 5 phrasings ($\geq$0) & 66.0\% \\
Docs improved or tied on majority of phrasings ($\geq$3/5) & 73.7\% \\
Docs improved or tied on all 5 phrasings ($\geq$0) & 22.2\% \\
Average gain when QI-GEO helps & +0.1634 \\
Average loss when QI-GEO hurts & $-$0.0629 \\
WTR (avg fraction of phrasings improved per doc) & 0.67 \\
DR (downside risk per doc, averaged) & 0.0097 \\
\bottomrule
\end{tabular}
\vspace{2mm}
\caption{Per-document multi-query stability metrics.}
\label{tab:mq_stability}
\end{table}

QI-GEO achieves a positive mean gain across all five query variants for 66.0\% of documents, indicating that the visibility improvements generally persist beyond a single query formulation. When considering majority agreement, 73.7\% of documents improve or maintain visibility on at least three of the five query variants. Furthermore, 22.2\% of documents improve across all five variants.

The gains are also substantially larger than the losses. When QI-GEO improves visibility, the average gain is +0.1634, compared to an average loss of -0.0629 when optimization is unsuccessful. This asymmetry is reflected in the low downside risk of 0.0097, suggesting that negative outcomes are relatively limited in magnitude.

\begin{figure}[H]
\centering
\includegraphics[width=1\linewidth, height = 4.3cm]{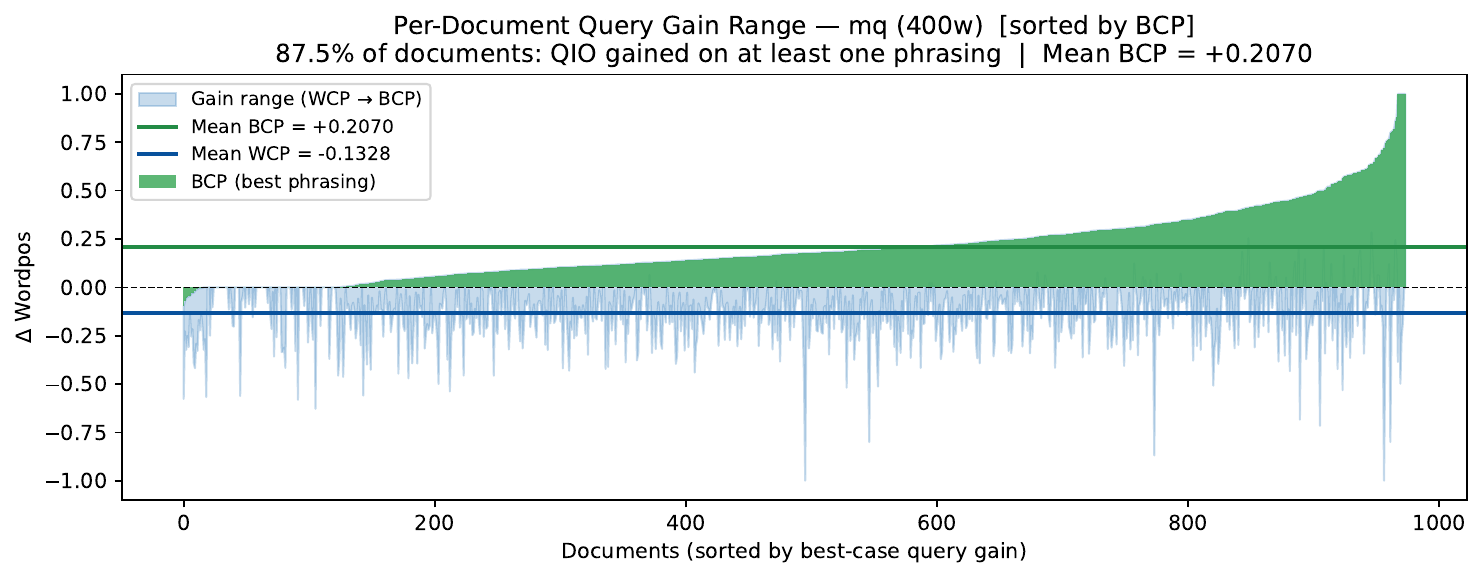}
\caption{Per-Document QI-GEO Gain Range across Query Phrasings}
\label{fig:qio_mq_bcp_mq}
\end{figure}

Figure~\ref{fig:qio_mq_bcp_mq} visualizes the range of gains observed across query variants for each document. While the exact improvement varies across formulations, the majority of documents exhibit at least one query variant for which QI-GEO provides a substantial benefit. In particular, 87.5\% of documents gain visibility on at least one phrasing of the same information need.

Overall, these results suggest that the visibility gains produced by QI-GEO are not restricted to a single query formulation. The improvements remain effective across a diverse set of semantically related queries, indicating that the document-centric intent representation captures information needs that generalize beyond individual query wording.

\section{Conclusion and Future Work}

This paper introduced Query-Implied Optimization (QI-GEO), a query-independent framework for Generative Engine Optimization. Existing GEO methods typically rely on explicit queries or query-derived signals to guide document optimization. In contrast, QI-GEO models user intent directly from document semantics. By extracting entities and relationships, expanding them through a knowledge graph, and identifying informational gaps, QI-GEO discovers concepts that may be relevant to user information needs without requiring access to future queries.

Experiments on GEO-Bench and ExtendedGEOBench show that QI-GEO improves citation coverage, source visibility, and subjective impression scores compared to the original documents. The results further demonstrate that these improvements remain effective across semantically related query variations, suggesting that document-centric intent modeling can serve as a viable alternative to query-based optimization strategies.

Several directions remain for future work. First, richer knowledge graph expansion strategies could improve the quality and coverage of discovered informational gaps. Second, future systems could jointly model multiple inferred intents during optimization rather than treating candidate gaps independently. Finally, while this work uses a general-purpose knowledge graph, domain-specific knowledge graphs may provide more precise semantic relationships and enable stronger optimization in specialized domains such as law, finance, healthcare, and geospatial information. These directions offer promising opportunities for improving the effectiveness and generality of query-independent GEO systems.

\bibliographystyle{IEEEtran}
\bibliography{references}

\end{document}